\documentclass[prl,groupedaddress,twocolumn]{revtex4}
\usepackage[dvips]{graphicx}

\begin{document}

\bibliographystyle{apsrev}

\title{Hot exciton transport in ZnSe quantum wells}

\author{Hui Zhao}
\author{Sebastian Moehl}
\author{Sven Wachter}
\author{Heinz Kalt}

\affiliation{Institut f\"{u}r Angewandte Physik, Universit\"{a}t Karlsruhe, D-76128 Karlsruhe, Germany}

\begin{abstract}

The in--plane transport of excitons in ZnSe quantum wells is
investigated directly by microphotoluminescence in combination
with a solid immersion lens. Due to the strong Fr\"ohlich
coupling, the initial kinetic energy of the excitons is well
controlled by choosing the excess energy of the excitation laser.
When increasing the laser excess energy, we find a general trend
of increasing transport length and more importantly a pronounced
periodic quenching of the transport length when the excess energy
corresponds to multiples of the LO--phonon energy. Such features
show the dominant role of the kinetic energy of excitons in the
transport process. Together with the excitation intensity
dependence of the transport length, we distinguish the phonon wind
driven transport of cold excitons and defect--limited hot exciton
transport.

\end{abstract}

\maketitle

The in-plane transport of excitons in semiconductor quantum wells
(QWs) has attracted a lot of interest due to both fundamental and
technological reasons. Generally, there are several possible
transport processes after an optical excitation. The first one is
the classical diffusion of cold excitons\cite{b3910901}. In this
case, the excitons have the same temperature as the lattice. The
transport can be well described by the diffusion equation, with a
constant diffusivity. The second process is the transport of hot
excitons\cite{b4613461}, initially ballistic (before the first
scattering event) and then diffusive. Since the excitons remain
hot, the transport is coupled with the relaxation process. This
kind of transport cannot be described by the diffusion equation,
since the effective 'diffusivity' varies in both temporal and
spatial domains. Beside these 'active' transport processes, which
are governed by the velocity of the excitons, the excitons can
also be passively driven by other factors, i.e., phonon
wind\cite{b391862}. Due to the increasing importance of
nanostructures, transport of excitons or carriers has to be
understood on a length scale comparable to the mean free path of
the particles. It is obvious that here strong deviations from
classical transport can be expected.

During the past two decades, the transport experiments have
concentrated mainly on III--V semiconductor QWs. The employed
optical techniques, namely
transient--grating\cite{b307346,b476827},
pump--probe\cite{b391862, b4613461},
time--of--flight\cite{b3910901,b4914523}, microphotoluminescence
($\mu$--PL)\cite{b584728} and near--field
pump--probe\cite{b602101}, have achieved an increasing spatial
resolution. The in--plane transport of excitons in II--VI QWs was
less investigated, and the transport has been regarded as
classical diffusion\cite{jap81536, apl74741}.

In this letter, we investigate the exciton in--plane transport in
ZnSe QWs on the length scale of few $\mu$m by solid immersion lens
(SIL)--enhanced $\mu$-PL. We show that the exciton transport in
this regime is not classical diffusion, but dominated by hot
exciton transport. In particular, we exploit the effect that, in
contrast to GaAs QWs, the initial kinetic energy of the excitons
in ZnSe QWs can be tuned in a well defined manner, due to
LO-phonon assisted generation of the excitons.

The confocal $\mu$--PL system consists of a microscope objective
(20$\times$, NA=0.4) and a hemisphere SIL\cite{apl741791} of
refractive index $n_{r}=2.2$. The SIL is adhesively fixed to the
sample surface. The spatial resolution of the whole system was
determined to be about 400 nm (FWHM of the Airy pattern). The
excitation source is a cw Ti:sapphire laser pumped by an Ar--ion
laser and frequency--doubled using a BBO crystal. The laser beam
is focused on the sample surface by the objective. The
luminescence is collected by the same objective. A 20 $\mu$m
pinhole in the image plane of the objective limits the detection
area to 455 nm in diameter. All measurements were performed at 7
K.
\begin{figure}[b]
 \includegraphics[width=8cm]{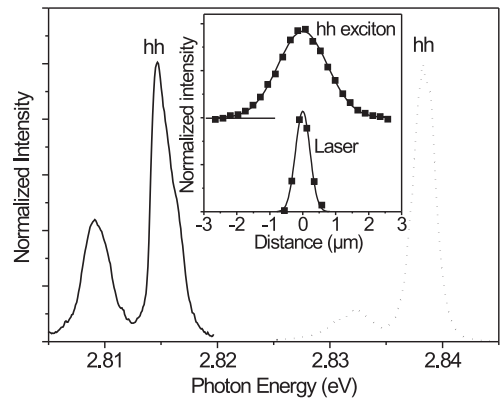}
 \caption{Photoluminescence of sample 1 (solid) and sample 2
(dots). Inset: An example of the spatial profiles of the
heavy--hole exciton luminescence (upper
 squares) and the laser spot (lower squares). The curves represent
 the corresponding Gaussian fits.
 }
 \label{fig1}
\end{figure}
Two samples are studied in this investigation. Sample 1 is a 120
periods of $\mathrm{ZnSe~(7.3~nm)/ZnS_{0.1}Se_{0.9}~(10.7~nm)}$
multiple QW grown by MOVPE. Sample 2 is a ZnSe~(5~nm) single QW
with $\mathrm{Zn_{0.9}Mg_{0.1}S_{0.16}Se_{0.84}}$ barriers grown
by MBE. The PL spectra of the two samples (Fig.~1) are dominated
by the peaks of the heavy hole excitons (hh), with similar
linewidth of about 2~meV. However, we note that the linewidth
cannot be used for comparing the quality of the two samples, since
the carrier confinements are different due to the different
barrier materials. Actually, the luminescence efficiency of sample
2 is two orders higher than that of sample 1, that indicates the
higher quality of sample 2. The samples were excited locally
through the objective, and the spectra from different positions of
the sample were measured by scanning the pinhole in the image
plane of the objective. This allows us to gain the spatial profile
of luminescence by drawing the spectrally integrated hh intensity
as function of spatial position. The upper squares in the inset of
Fig.~1 give an example of the spatial profiles measured in this
way. By Gaussian fit, shown as the upper curve, we obtained the
transport length as the FWHM of the Gaussian function. The profile
of the laser spot was also measured during the same scan, as shown
in the lower part of the inset.

The exciton formation and relaxation processes in these samples
are well understood from monitoring the temporal evolution of the
LO--phonon replica\cite{b571390}. In Fig.~2, we show a schematic
\begin{figure}[h]
 \includegraphics[width=6cm]{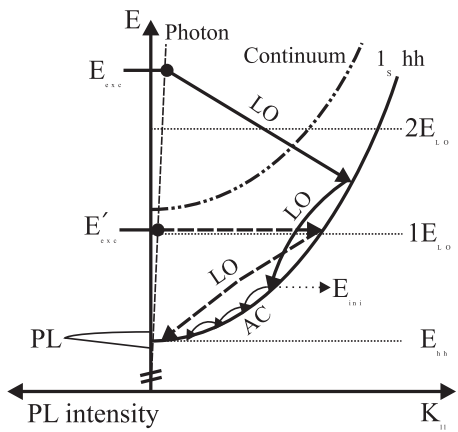}
 \caption{ Schematic drawing of the exciton formation assisted by
LO--phonon emission and the subsequent relaxation by LO--phonon
and acoustic phonon (AC) emissions.
 }
 \label{fig1}
\end{figure}
drawing of these processes. After the optical excitation, the
electron--hole pairs created in continuum states form excitons
directly assisted by emission of LO--phonons within few ps, due to
the strong Fr\"ohlich coupling in polar II-VI quantum structures.
These hot excitons relax rapidly by emission of LO--phonons, until
their kinetic energy is lower than the LO--phonon energy, $E_{LO}$
(31.8 meV for both samples, measured by Raman spectroscopy). Then
the relaxation can only be achieved slowly by emission of acoustic
phonons and continues over some 100 ps\cite{b571390}. We define
the excess energy of the excitation, $E_{excess}$, as the
difference between the laser energy and the energy of the hh
exciton resonance, $E_{laser}-E_{hh}$. In this investigation,
$E_{excess}$ is chosen such that the exciton can only emit one or
two LO-phonons after formation. In this case, we can define the
exciton initial kinetic energy of the slow relaxation process
realized by acoustic phonon emission to be
$E_{ini}=E_{excess}-nE_{LO}$, in which $n$ equals to 1, 2 or 3
depending on $E_{excess}$. By tuning the $E_{excess}$, we can
periodically tune the $E_{ini}$, and thus the initial in--plane
group velocity of the excitons, to investigate the influence of
this velocity on the transport process.

Figure 3 shows the excess energy dependence of the transport
length for both samples. When $E_{excess}<E_{LO}$, the
luminescence is very weak due to the inefficient exciton formation
since the LO--phonon emission path is not available. We note that
in the case of $E_{excess} \approx E_{LO}$, the hh peak is
\begin{figure}[h]
 \includegraphics[width=8cm]{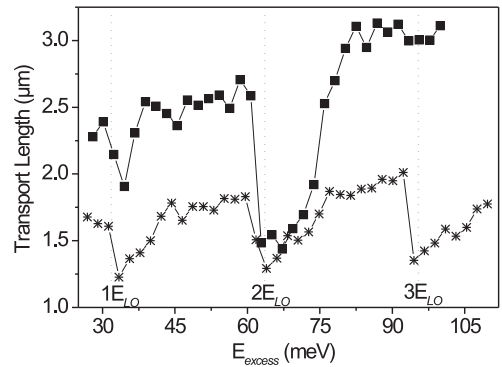}
 \caption{Excess energy dependence of the transport length at an
excitation intensity of 1 $\mathrm{kW/cm^{2}}$. Sample 1: stars;
sample 2: squares.
 }
 \label{fig1}
\end{figure}
superimposed by a strong resonant Raman scattering peak. By
multi-peak fitting, we can subtract the later from the hh peak, so
the spatial profile gained in this case is not influenced by the
Ramam scattering. In Fig.~3, we can observe two features: (1) a
general trend of increasing transport length with increasing
$E_{excess}$ and more importantly (2) a pronounced periodic
quenching of the transport length when $E_{excess} \approx
nE_{LO}$.

In Fig.~4, we show the excitation intensity dependence of the
transport length of sample 1. Several values of $E_{excess}$ are
chosen in the range of $1E_{LO}$ to $2E_{LO}$, and the
corresponding values of the $E_{ini}$ are shown in Fig.~4. We
estimate the exciton density from the excitation intensity, as
shown in Fig.~4 as the top axis. An absorption coefficient of
$6.5\times10^{4}/\mathrm{cm}$ (measured by absorption
spectroscopy) and a decay time of 300 ps (measured by
time-resolved photoluminescence) are used for this estimation. We
find from Fig.~4 that in the cases of small $E_{ini}$ (1 meV and 7
\begin{figure}[h]
 \includegraphics[width=8cm]{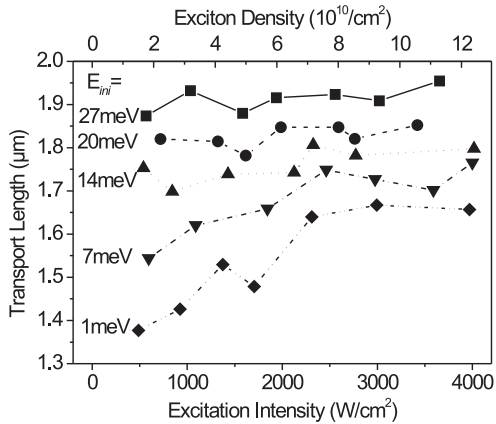}
 \caption{Excitation intensity dependence of the transport length
of sample 1.
 }
 \label{fig1}
\end{figure}
meV), the transport length increases sub--linearly with the
excitation intensity. For higher $E_{ini}$, the transport length
is independent on the excitation intensity.

In the case of phonon wind driven transport, the transport length
increases with the excitation intensity\cite{b391862}, while for
the classical diffusion of cold excitons and the hot exciton
transport, the transport lengths are independent on the excitation
intensity. In our experiment, when the $E_{excess}$ is slightly
larger than $nE_{LO}$, cold excitons are generated with small
$E_{ini}$. According to the excitation intensity dependence
behavior, these cold excitons are driven by phonon wind.
Increasing the $E_{excess}$ within one period, we observe the
increase of the transport length in Fig.~3. This behavior can be
explained by either the hot exciton transport or the phonon wind
driven model. In the former case the $E_{ini}$ increases with
$E_{excess}$, while in the latter case, the force of the phonon
wind also increases with the $E_{excess}$ due to the increasing of
the number of phonons emitted during the relaxation of the
excitons. However, the latter possibility can be ruled out by the
strong periodic feature observed in Fig.~3. As mentioned above,
the $E_{ini}$ is a periodic function of the $E_{excess}$ with a
period of $E_{LO}$ due to the fast LO-phonon emission. Thus, the
periodic feature can be well explained by the hot exciton
transport. In the phonon wind driven model, the wind force is
anticipated to be increase monotonously with $E_{excess}$, since
the LO-phonons emitted during the relaxation decay into acoustic
phonons within few ps\cite{b5014179}. For these reasons, we
attribute the transport of the excitons with high $E_{ini}$ to the
hot exciton transport. Actually, since the phonon wind cannot
drive the excitons at drift velocities exceeding the sound
velocity\cite{b391862}, the influence of the phonon wind on the
hot excitons should be weak. The above explanations are confirmed
by the independence of transport length on the excitation
intensity for hot excitons, as observed in Fig.~4.

Finally, we discuss briefly the influence of defects on the
transport properties. In Fig.~3, we find the transport length of
sample 2 is larger than that of sample 1. The dip around $1E_{LO}$
is more pronounced in sample 1 than in sample 2, comparing with
the dips around $2E_{LO}$. In the case of $E_{excess} \approx
1E_{LO}$ (see $E^{'}_{exc}$ in Fig.~2), the exciton formation is
inefficient due to the small momentum difference. However, the
formation can be assisted by the relaxation of momentum
conservation due to defects (shown as the dashed line after the
$E^{'}_{exc}$ in Fig.~2). Thus, the dip around 1$E_{LO}$ is
anticipated to be more pronounced in the sample containing more
defects (sample 1, due to the smaller transport length and the
lower luminescence efficiency). Furthermore, the periodic feature
of the transport length is more pronounced in sample 1 than in
sample 2 (see the dips around 3$E_{LO}$). This coincides with the
fact that normally in PLE, the LO--phonon cascades are easier to
be observed in the samples containing more defects, and can be
attributed to the loss of excitons during the slow relaxation
process. The difference of the two samples implies that the hot
exciton transport in these samples is limited by defects. This
statement is also consistent with calculations of the transport
length from exciton acoustic-phonon interaction neglecting
defects, which yields expected values of about 10 $\mu$m.

In summary, we measured directly the in-plane transport of
excitons in ZnSe QWs on the length scale of few $\mu$m by
SIL--enhanced $\mu$--PL. Due to the strong Fr\"ohlich coupling,
the initial kinetic energy of the excitons is well controlled by
tuning the laser excess energy. We find the dominant role of the
kinetic energy of excitons in the transport process. The cold
excitons are driven by the phonon wind. For the excitons with high
initial kinetic energies, the process is dominated by the hot
exciton transport. Furthermore, the hot exciton transport is
limited by defects.\newline

We gratefully acknowledge the growth of the excellent samples by
the group of M. Heuken (Aachen) and the group of D. Hommel
(Bremen). This work was supported by the Deutsche
Forschungsgemeinschaft.

\end{document}